\documentclass{aa}  

\usepackage{graphicx}
\usepackage{txfonts}
\usepackage{lipsum}
\usepackage{subcaption}         
\usepackage{lscape}             
\usepackage{placeins}           

\usepackage{breqn}              

\usepackage{xcolor}

\usepackage[
  colorlinks=true,
  linkcolor=blue,
  citecolor=blue,
  urlcolor=blue
]{hyperref}

\newcommand{\Matrix}[1]{\mathbf{\hat{#1}}}
\newcommand{\Vector}[1]{\vec{#1}}

\newcommand{\GADGETFOUR}{\texttt{GADGET4}\xspace}
\newcommand{\monofonic}{\texttt{monofonIC}\xspace}

\newcommand{\CAMB}{\texttt{CAMB}\xspace}

\newcommand{\nbodykit}{\texttt{nbodykit}\xspace}
\newcommand{\crispy}{\texttt{CRISPy}\xspace}

\newcommand{\HMCODE}{\texttt{HMCODE}\xspace}

\newcommand{\lcdm}{$\Lambda$CDM\xspace}

\newcommand{\Tcube}{$\mathrm{T}^3$\xspace}
\newcommand{\Rcube}{$\mathbb{R}^3$\xspace}

\newcommand{\Oh}{\mathrm{O_h}}
\newcommand{\SOthree}{\mathrm{SO(3)}}

\newcommand{\Nbody}{$N$-body\xspace}
\newcommand{\SUBFIND}{\texttt{SUBFIND}\xspace}

\newcommand{\kpc}{\mathrm{kpc}}
\newcommand{\mpc}{\mathrm{Mpc}}

\newcommand{\msol}{\mathrm{M_{\odot}}}

\begin{document}

   \title{The box within the Universe: anisotropies in periodic cosmological simulations}

   \subtitle{I. Dark-matter-only simulations}

%
%
%

   \author{G. R\'acz\corrauth{gabor.racz@helsinki.fi}        
        \and T. Sawala
        }

   \institute{Department of Physics, University of Helsinki, Gustaf H\"allstr\"omin katu 2, FI-00014 Helsinki, Finland}

   \date{Received November 4, 2026}

 
  \abstract
   {The vast majority of cosmological simulations employ periodic boundary conditions to achieve statistical homogeneity and to mimic the infinite spatial extent of our Universe. In a cubic periodic domain the Green function breaks full rotational invariance and introduces a geometry-induced anisotropy in the gravitational interaction.}
   {We quantify how this intrinsic anisotropy affects cosmic structure growth and investigate the extent to which it biases the commonly used large-scale structure and halo-based observables in dark-matter-only simulations.} 
   {We analyse an ensemble of 498 cosmological dark-matter-only \Nbody simulations in cubic \Tcube topological manifolds, evolved to $z=0$, spanning box sizes from $L=6.25\,\mathrm{Mpc}$ to $L=100\,\mathrm{Mpc}$ at fixed mass and force resolution. We compare anisotropy-sensitive statistics across the suite, including the matter power spectrum, halo abundance, merger histories, shapes and alignments, and filament orientations.}
   {Finite periodic boundaries imprint orientation-dependent signatures aligned with the box geometry. These signatures propagate beyond the largest modes and can induce systematic shifts in both field-level and halo-level statistics. The detectable effects propagate to scales much smaller than the box size.}
   {Our results demonstrate that periodic-box anisotropy is a non-negligible source of systematics for small simulation volumes, including those commonly employed in modern hydrodynamical simulations.}


   \maketitle
   \nolinenumbers

\section{Introduction}
Simulations of non-linear cosmological structure formation are the primary tools for studying the evolution and internal structure of dark matter haloes, galaxies, galaxy groups and clusters, as well as the cosmic web. These simulations follow the coupled evolution of matter components under gravity in an expanding background. The background expansion is set by the adopted cosmological model, most commonly the standard \lcdm scenario, in which the energy density at late times is dominated by a cosmological constant ($\Lambda$) and collisionless cold dark matter (CDM), with a subdominant baryonic component and, in many applications, additional species such as massive neutrinos. In dark-matter-only \Nbody simulations, the matter content is represented by collisionless tracer particles sampling the underlying density and velocity fields, and the evolution is driven solely by gravity \citep{2022LRCA....8....1A}. Hydrodynamical simulations extend this framework by including gas dynamics and subgrid prescriptions for radiative cooling, star formation, chemical enrichment, and feedback processes \citep{2023ARA&A..61..473C}.

The vast majority of cosmological simulations are performed in cubic volumes with periodic boundary conditions, which mimic an infinite universe while keeping the computational domain finite \citep{1985ApJS...57..241E, 1991ApJS...75..231H}. Mathematically, these boundary conditions correspond to evolving the system on a three-torus, which we denote by the cubic \Tcube manifold, topologically equivalent to a Cartesian product of three circles ($\mathrm{S}^1 \times \mathrm{S}^1 \times \mathrm{S}^1$). While a non-trivial global topology is not excluded in principle \citep{1984PAZh...10..323Z, 1997Natur.385..139E}, current large-scale structure and CMB analyses strongly constrain simple compact topologies on scales comparable to the observable Universe \citep{2002MNRAS.336L..13H, 2016Univ....2....1L}. In practice, the \Tcube periodicity is adopted primarily as a numerical convenience: in a periodic domain, the Poisson equation admits efficient Fourier-space solutions, enabling particle-mesh (PM) schemes and hybrid methods such as TreePM, which combines a Fast Fourier Transform (FFT)-based long-range force calculation with a high-resolution short-range solver \citep{1983MNRAS.204..891K, 1995ApJS...98..355X}. Periodic boundaries also eliminate physical edges and the associated boundary artefacts that would otherwise contaminate the global statistics of the simulation volume.
Despite these advantages, a periodic cubic domain is not fully equivalent to an infinite Euclidean space \Rcube. In particular, while the periodic domain preserves continuous translational invariance, it does not preserve full rotational invariance: the periodic Green function of the Poisson equation in a cube carries an imprint of the box geometry. As a result, the effective gravitational interaction in \Tcube can be viewed as the Newtonian force supplemented by contributions from the infinite set of periodic images, which collectively generate a tidal field tied to the lattice directions. This induces a geometry-driven anisotropy that can bias anisotropy-sensitive observables, including halo shapes and internal 3D profiles, orientation and alignment statistics, angular multipoles of the density field, and filament orientations. From a symmetry perspective, the cubic periodic volume is not invariant under the full rotation group $\SOthree$, but only under its discrete octahedral subgroup $\mathrm{O}$. Including reflections and inversion, its point symmetries form the full octahedral group $\Oh$.
Consequently, the Lagrangian of a self-gravitating system evolved in a cubic periodic domain inherits the symmetries of the simulation volume rather than those of isotropic Euclidean space. In particular, because continuous rotational invariance is broken to a discrete subgroup, Noether’s theorem no longer implies conservation of the total angular momentum of the system, even in the absence of external torques. By contrast, periodic boundary conditions preserve continuous translational invariance, and therefore the corresponding total linear momentum is conserved by the equations of motion.
Importantly, these effects are conceptually distinct from the well-known finite-volume systematics associated with missing long-wavelength modes \citep{BaglaPrasad2006, Orban2014, 2011ApJS..194...46G, 2016JCAP...09..007B}: truncating fluctuations with wavelengths larger than the box affects growth through the absence of large-scale density and tidal fields, whereas periodic-force anisotropy reflects the symmetry of the Green function itself in a cubic fundamental domain. In the non-linear regime, mode coupling can further propagate such anisotropic signatures to scales substantially smaller than the box size, complicating the interpretation of results from small-volume simulations.

Small periodic volumes are now routine in high-resolution galaxy-formation simulations, where computational cost forces a trade-off between resolution and volume. Recent examples include EAGLE \citep{2015MNRAS.446..521S}, which provides standard runs with $L=25$--$100\,\mpc$, SIMBA \citep{2019MNRAS.486.2827D} with $L=18.4$--$147.1\,\mpc$, IllustrisTNG \citep{2019MNRAS.490.3196P} with $L=50,100,300\,\mpc$, CROCODILE \citep{2024ApJ...975..183O} with $L=25,50\,\mpc$, and COLIBRE \citep{2026MNRAS.548ag375S} spanning $L=25$--$400\,\mpc$.
Recent work has also identified boundary-crossing artefacts whose prevalence increases towards smaller simulation volumes, with the smallest TNG boxes being most affected \citep{2025A&A...703L..16M}. These artefacts are distinct from the intrinsic anisotropy of the periodic gravitational interaction studied here.

\citet{2021MNRAS.503.5638R} demonstrated that cubic periodicity introduces anisotropy in the matter power spectrum. Here, we extend this investigation to halo properties and assembly histories, and to the orientations of haloes and filaments, with particular relevance to the small volumes used in high-resolution hydrodynamical simulations. This paper is part of a series that explores the effects of box-induced anisotropies in cosmological simulations:
\begin{enumerate}
    \renewcommand{\theenumi}{\Roman{enumi}}
    \item Dark-matter-only simulations (this work)
    \item Hydrodynamical simulations (Sawala et al. in prep.)
\end{enumerate}

Since gravity also drives structure formation in hydrodynamical simulations, the goal of this paper is to explore the leading-order signatures of periodic-box anisotropy, which we expect to already be present in the collisionless dark matter distribution.

\section{Newtonian gravity in a periodic box}

\subsection{Symmetries of a periodic cubic volume}
\label{sec:symmetries}

\begin{figure}[h!]
    \centering
    \begin{subfigure}[a]{0.495\textwidth}
        \centering
        \includegraphics[width=0.99\textwidth]{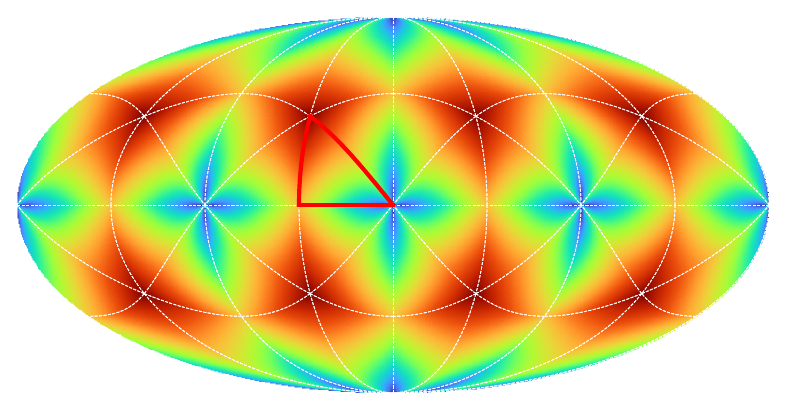}
        \caption{Tiled fundamental wedge over the full sphere}
    \end{subfigure}
    ~
    \centering
    \begin{subfigure}[b]{0.495\textwidth}
        \centering
        \includegraphics[width=0.99\textwidth]{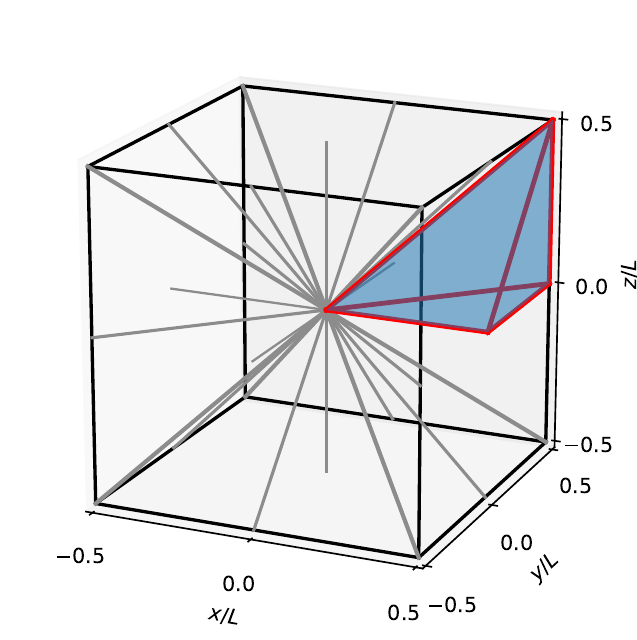}
        \caption{Cubic periodic volume and the fundamental tetrahedron}
    \end{subfigure}

    \caption{Symmetries of a cubic periodic volume. \textbf{Top:} Mollweide visualisation of the cube’s octahedral point-group symmetry. The coloured pattern is defined on a single fundamental $1/48$ spherical wedge and then replicated over the full sphere by the $\Oh$ symmetry operations. Thin white curves indicate the full set of wedge boundaries, while the red curve outlines the reference fundamental wedge. \textbf{Bottom:} Three-dimensional view of the periodic cube with symmetry axes shown as grey lines. The highlighted fundamental tetrahedron (blue faces with red edges) corresponds to the region $x \ge y \ge z \ge0$ in minimum-image coordinates and has volume $L^3/48$.}
    \label{fig:FundamentalTetrahedronAndWedge}
\end{figure}

We consider a cubic simulation volume of side length $L$ with periodic boundary conditions. The spatial manifold is the three-torus \Tcube obtained by connecting the opposite faces of the cube. Periodicity implies invariance under continuous translations on \Tcube just as in \Rcube, so there is no physically preferred origin or box centre from the perspective of an observer living in \Tcube. The periodic box is not invariant under the full rotation group $\SOthree$, but only under its discrete octahedral subgroup, $\mathrm{O}$. Including reflections and inversions, its point symmetries form the full octahedral group $\Oh$ \citep{2016gtnp.book.....Z}. This reduction of symmetry is important to the orientation-dependent systematics studied in this work, and it also provides a practical tool: we can fold anisotropy-sensitive observables into the fundamental domain of $\Oh$ to increase their signal-to-noise ratio \citep{2021MNRAS.503.5638R}.

The three-torus can be written as
\begin{equation}
  \mathrm{T}^3 \;\simeq\; \mathbb{R}^3/(L\mathbb{Z})^3 ,
\end{equation}
so translations on \Tcube form a continuous group. For any displacement $\Vector{a}\in\mathbb{R}^3$, the map
\begin{equation}
  [\Vector{x}] \mapsto [\Vector{x}+\Vector{a}]
\end{equation}
is a well-defined isometry of \Tcube, where $[\Vector{x}]$ denotes the equivalence class modulo $(L\mathbb{Z})^3$.
The covering space is $\mathbb{R}^3$; each point on the torus can be represented by coordinates in a chosen fundamental cube, e.g.\ $\Vector{x}\in[0,L)^3$.
In this representation, points related by
\begin{equation}
  \Vector{x} \sim \Vector{x} + L\Vector{n}, \qquad \Vector{n}\in\mathbb{Z}^3,
  \label{eq:lattice_identifications}
\end{equation}
correspond to the same point on \Tcube. The discrete lattice $(L\mathbb{Z})^3$ therefore appears as the set of ``image labels'' used for wrapping positions and defining minimum-image separations.

While translations are continuous, rotational invariance is reduced by the cubic geometry. Elements of $\Oh$ can be represented by signed permutation matrices $\Matrix{R}$. In these matrices, each row and each column contains exactly one non-zero element equal to $\pm1$. There are $48$ such matrices, corresponding to the $48$ elements of $\Oh$. The subset with $\det(\Matrix{R})=+1$ forms the rotational octahedral group $\mathrm{O}$ with $24$ elements, while the remaining $24$ elements with $\det(\Matrix{R})=-1$ include reflections and inversion combined with rotations \citep{1991AlbertCotton}.

Let $g=(\Vector{a},\Matrix{R})$ denote an isometry composed of a translation on \Tcube by $\Vector{a}$ and a point-group operation $\Matrix{R}\in O_h$. Its action on positions is
\begin{equation}
  [\Vector{x}] \mapsto [\Matrix{R}\Vector{x} + \Vector{a}] .
\end{equation}
The induced action on fields follows standard tensorial rules. For a scalar field $s(\Vector{x})$ within the periodic box (e.g. density),
\begin{equation}
  s(\Vector{x}) \mapsto s'(\Vector{x}) = s(g^{-1}\Vector{x}).
  \label{eq:scalar_transform}
\end{equation}
For a vector field $\Vector{v}(\Vector{x})$ (e.g. velocity),
\begin{equation}
  \Vector{v}(\Vector{x}) \mapsto \Vector{v}'(\Vector{x})
  = \Matrix{R}\,\Vector{v}(g^{-1}\Vector{x}) .
  \label{eq:vector_transform}
\end{equation}
For a rank-2 tensor field $\Matrix{T}(\Vector{x})$ (e.g. inertia or tidal tensors),
\begin{equation}
  \Matrix{T}(\Vector{x}) \mapsto \Matrix{T}'(\Vector{x})
  = \Matrix{R}\,\Matrix{T}(g^{-1}\Vector{x})\,\Matrix{R}^{\mathrm{T}} .
  \label{eq:tensor_transform}
\end{equation}

Because the cube's point symmetry is discrete, one may restrict attention to a fundamental region of the simulation volume defined by $\Oh$. This is useful for folding measurements that depend on orientation relative to the box geometry. A convenient choice is obtained by working with coordinates relative to a pre-selected box centre. Note that, due to translation invariance on \Tcube, this choice of centre is free. In practice, we work with relative positions, either with respect to a fixed reference point or with respect to an object of interest (e.g. a halo centre).

We define the minimum-image relative coordinate
\begin{equation}
  \tilde{\Vector{x}} \equiv \mathrm{wrap}\!\left(\Vector{x}-\Vector{x}_0\right) \in \left[-\tfrac{L}{2},\tfrac{L}{2}\right)^3,
\end{equation}
where $\Vector{x}_0$ is the chosen reference point and $\mathrm{wrap}$ applies the periodic identifications of Eq.~\ref{eq:lattice_identifications}. Using reflections, we map to the first octant, and using axis permutations we order the components. This yields a representative (unique up to measure-zero boundary degeneracies) in the region
\begin{equation}
  \mathcal{F} \equiv
  \left\{
  \tilde{\Vector{x}} \in \left[0,\tfrac{L}{2}\right]^3 :
  \tilde{x}_1 \ge \tilde{x}_2 \ge \tilde{x}_3 \ge 0
  \right\}.
  \label{eq:fundamental_tetrahedron}
\end{equation}
The domain $\mathcal{F}$ is a tetrahedron with vertices
\begin{equation}
  (0,0,0),\quad \left(\tfrac{L}{2},0,0\right),\quad
  \left(\tfrac{L}{2},\tfrac{L}{2},0\right),\quad
  \left(\tfrac{L}{2},\tfrac{L}{2},\tfrac{L}{2}\right),
\end{equation}
and volume $V_{\mathcal{F}} = L^3/48$, consistent with the $48$ elements of $\Oh$. Geometrically, $\mathcal{F}$ selects one of the $48$ congruent tetrahedra into which the cube can be partitioned by the $\Oh$ symmetry transformations. 

Many of our observables depend only on directions (e.g. halo major axes, filament orientations) rather than positions. For a unit vector $\hat{\Vector{n}}$, a convenient direction-space fundamental domain is obtained by applying sign flips and permutations until
\begin{equation}
  \hat{n}_1 \ge \hat{n}_2 \ge \hat{n}_3 \ge 0 .
  \label{eq:fundamental_wedge}
\end{equation}
This defines a $1/48$ wedge of the unit sphere in cubic \Tcube space. The fundamental tetrahedron and fundamental wedge are visualised in Fig.~\ref{fig:FundamentalTetrahedronAndWedge}.

Using the fundamental tetrahedron and fundamental wedge, we can increase the signal-to-noise ratio of anisotropy-sensitive statistics. For example, if a filament or halo major axis points towards the diagonal of the cube, it is unimportant which of the eight corners it points to; likewise, alignments towards face normals or edge directions are defined only up to the cube symmetry. For each direction-dependent measurement, we can construct a point-group operation $\Matrix{R}\in \Oh$ that maps the relevant direction -- and, if needed, the relevant relative position -- into the chosen fundamental domain, and we rotate associated quantities using Eqs.~\ref{eq:scalar_transform}--\ref{eq:vector_transform}. We then accumulate statistics in $\mathcal{F}$ or in the fundamental direction wedge. This procedure is effectively an average over equivalent regions under the $\Oh$ symmetries of the cube, improving the precision of symmetry-averaged anisotropy statistics while preserving genuine box-aligned anisotropy.

\subsection{The Green function}

\begin{figure*}[ht!]
    \centering
    {\includegraphics[width=0.99\textwidth]{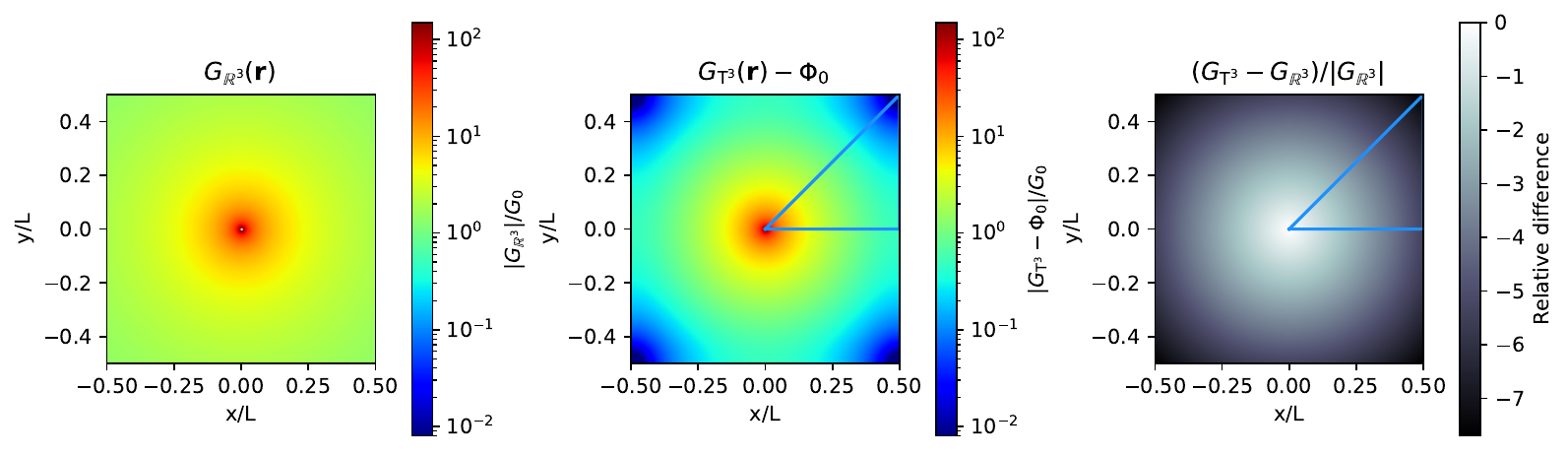}}
     \caption{The Green function of Poisson's equation for gravity in different topological manifolds. \textbf{Left:} The Green function in trivial \Rcube topology. This function is isotropic and $G_{\mathbb{R}^3}(|\Vector{r}|) \propto |\Vector{r}|^{-1}$ which is equivalent to Newton's law of universal gravitation. \textbf{Centre:} The zero-mean Green function on the fully periodic cubic space \Tcube, shown in the $z=0$ plane after subtracting an additive matching constant $\Phi_0$. The periodic images make the field anisotropic, while retaining the discrete cubic symmetries of the box. \textbf{Right:} The relative difference, $(G_{\mathrm{T}^3}-G_{\mathbb{R}^3})/|G_{\mathbb{R}^3}|$, between the \Tcube and \Rcube Green functions. The first two panels are shown in units of $G_0=1/(4\pi L)$, where $L$ is the box size.}
      \label{fig:PeriodicGreen}
\end{figure*}

During cosmic structure formation, the metric perturbations remain small and the matter field is well described as a slow-moving, collisionless fluid on sub-horizon scales. In this regime, the Newtonian limit is sufficient to describe gravitational interactions. The ``peculiar'' gravitational potential $\Phi(\mathbf{x})$ is the solution of Poisson's equation
\begin{equation}
    \nabla^2 \Phi(\Vector{x}) = 4\pi G\,\left[\rho(\Vector{x})-\bar{\rho}\right],
    \label{eq:Poisson}
\end{equation}
where $G$ is Newton's constant, $\Vector{x}$ is the comoving position, $\rho(\mathbf{x})$ is the comoving mass density, and $\bar{\rho}$ is its mean. We use the rescaled peculiar potential $\Phi=a\phi$, where $\phi$ is the conventional peculiar gravitational potential and $a$ is the cosmic scale factor \citep{2005MNRAS.364.1105S}. The Newtonian Green function $G(\Vector{x},\Vector{x}')$ is defined as the fundamental solution of the Poisson operator,

\begin{equation}
    \nabla^2 G(\Vector{x},\Vector{x}') = \delta^{(3)}(\Vector{x}-\Vector{x}') ,
    \label{eq:GreenFunction}
\end{equation}
By superposition, the potential can be written as the convolution
\begin{equation}
    \Phi(\Vector{x}) = 4\pi G \int_V G(\Vector{x},\Vector{x}')\,\left[\rho(\Vector{x}')-\bar{\rho}\right]\,\mathrm{d}^3x' ,
    \label{eq:Phi_convolution}
\end{equation}
where $V$ denotes the simulation volume, and $\delta^{(3)}$ is the 3D Dirac-delta function. The Green function depends on the spatial domain and its boundary conditions. In \Rcube, imposing vanishing potential at infinity gives
\begin{equation}
    G_{\mathbb{R}^3}(\Vector{x},\Vector{x}') = -\frac{1}{4\pi \left| \Vector{x} - \Vector{x}' \right|}.
    \label{eq:RcubeGreen}
\end{equation}
Note that this Green function has the full rotational invariance of $\SOthree$.

In cubical \Tcube space, the Green function must satisfy periodicity in each coordinate. The periodic potential is generated by the source and all of its periodic images. Formally,
\begin{equation}
    G_{\mathrm{T}^3}(\Vector{x},\Vector{x}')
    \propto -\sum_{\Vector{n}\in\mathbb{Z}^3}\frac{1}{4\pi|\Vector{x}-\Vector{x}'-\Vector{n}L|},
    \label{eq:T3Green_images_formal}
\end{equation}
but this real-space lattice sum is only conditionally convergent and must be regularised with a uniform compensating background that removes the mean source density.

A convenient representation is obtained in Fourier space. Writing $\Vector{r}\equiv \Vector{x}-\Vector{x}'$ and $\Vector{k} = (2\pi/L)\,\Vector{m}$ with $\Vector{m}\in\mathbb{Z}^3$, the periodic Green function with vanishing mean can be written as
\begin{equation}
    G_{\mathrm{T}^3}(\Vector{r})
    = -\frac{1}{L^3}\sum_{\Vector{k}\neq \Vector{0}}
    \frac{e^{i\Vector{k}\cdot\Vector{r}}}{k^2}.
    \label{eq:T3Green_fourier}
\end{equation}
The exclusion of the $\Vector{k}=\Vector{0}$ mode enforces
\begin{equation}
    \int_V G_{\mathrm{T}^3}(\Vector{r})\,\mathrm{d}^3r = 0
\end{equation}
and corresponds to solving Poisson's equation for $\left[\rho(\Vector{x}')-\bar{\rho}\right]$ rather than $\rho(\Vector{x})$. Because $G_{\mathrm{T}^3}(\Vector{r})$ is constructed from a cubic periodic domain, it inherits the discrete $\Oh$ symmetries of the simulation volume. As a consequence, it is enough to calculate the periodic Green function within the fundamental tetrahedron of the cube, and the usual $\Oh$ symmetries can be used to tile the full periodic box. The non-periodic $G_{\mathbb{R}^3}(\Vector{r})$ and the periodic $G_{\mathrm{T}^3}(\Vector{r})$ Green functions are compared in Fig.~\ref{fig:PeriodicGreen}.
One interesting property of the periodic Green function is that $|\vec{\nabla} G_{\mathrm{T}^3}(\Vector{r})| \,< | \vec{\nabla} G_{\mathbb{R}^3}(\Vector{r}) |$ within the periodic cubic volume. As a consequence, the gravitational interactions are always weaker in a \Tcube manifold compared to an open \Rcube space.

\subsection{The Hamiltonian and the integrals of motion}

In comoving coordinates, the Hamiltonian
\begin{dmath}
    \mathcal{H}(\Vector{x}_1,\dots,\Vector{x}_N, \Vector{p}_1,\dots,\Vector{p}_N,t) = \sum\limits_{i}\frac{\Vector{p}^2_i}{2 m_i a(t)^2}+\frac{1}{2\,a(t)}\sum\limits_{i} m_i\,\Phi(\Vector{x}_i)
    \label{eq:hamiltonian}
\end{dmath}
 describes a self-gravitating particle system in an expanding background, where $\Vector{x}_i$ and $\Vector{p}_i$ are the comoving coordinate and canonical momentum, $m_i$ is the particle mass, and $a(t)$ is the cosmic scale factor. For a system of discrete particles, the rescaled peculiar potential $\Phi(\Vector{x}_i)$ can be expressed through the Green function as
\begin{equation}
    \Phi(\Vector{x}_i)=4\pi G \sum_{\substack{j=1\\ j\neq i}}^{N} m_j\,G(\Vector{x}_i-\Vector{x}_j),
    \label{eq:phi_from_green}
\end{equation}
with $G=G_{\mathbb{R}^3}$ for free-space interactions and $G=G_{\mathrm{T}^3}$ for cubic periodic interactions.
Since $\mathcal{H}$ depends explicitly on time through $a(t)$, 
\begin{equation}
    \frac{\mathrm{d}\mathcal{H}}{\mathrm{d}t}=\frac{\partial \mathcal{H}}{\partial t}\neq 0.
\end{equation}
As a consequence, the Hamiltonian itself is not a constant of motion.

Both $G_{\mathbb{R}^3}(\Vector{r})$ and $G_{\mathrm{T}^3}(\Vector{r})$ depend only on the separation vector $\Vector{r}=\Vector{x}_i-\Vector{x}_j$. Therefore $\mathcal{H}$ is invariant under uniform translations of all particle positions,
\begin{equation}
    \Vector{x}_i \rightarrow \Vector{x}_i + \Vector{\epsilon}
    \qquad \forall i
\end{equation}
in the \Rcube manifold, while
\begin{equation}
    [\Vector{x}_i] \rightarrow [\Vector{x}_i + \Vector{\epsilon}]
    \qquad \forall i
\end{equation}
in \Tcube. The generator of translations is the total canonical momentum
\begin{equation}
    \Vector{P}\equiv \sum_{i=1}^{N}\Vector{p}_i.
    \label{eq:total_momentum}
\end{equation}
In comoving coordinates, this corresponds to conservation of the appropriate comoving total momentum. A practical choice in \Nbody simulations is to set this to zero in the initial conditions, to avoid drifts of particles in the simulation volume.

In \Rcube, the Green function $G_{\mathbb{R}^3}(\Vector{r})=G_{\mathbb{R}^3}(|\Vector{r}|)$ is isotropic, and as a consequence, the full Hamiltonian is invariant under continuous $\SOthree$ rotations. The generator of these transformations is the total angular momentum
\begin{equation}
    \Vector{L}\equiv \sum_{i=1}^{N}\Vector{x}_i\times \Vector{p}_i
    \label{eq:total_angular_momentum}
\end{equation}
and one finds $\mathrm{d}\Vector{L}/\mathrm{d}t=\Vector{0}$ based on Noether's theorem \citep{1969mech.book.....L}. In contrast, on a \Tcube manifold, the periodic Green function $G_{\mathrm{T}^3}$ is invariant only under the discrete $\Oh$ symmetry group, but not under the full continuous rotation group $\SOthree$. Consequently, there is no continuous rotational symmetry and Noether's theorem does not imply conservation of $\Vector{L}$. This is a clear consequence of the fact that the periodic gravitational force is not strictly central,
because of the contribution of the periodic images. Therefore the net internal torque need not vanish, and the total angular momentum of the simulated matter distribution can in principle evolve with time.

While the conservation of the angular momentum plays an important role in structure formation, this quantity is not conserved in fully periodic cosmological simulations. The practical consequence is that orientation- and spin-related diagnostics can acquire subtle, box-aligned systematics. This is expected to be most relevant in small simulation volumes, where the characteristic size of non-linear structures is not negligible compared to the box size. In such cases, the box-induced torques can influence the time evolution of halo angular momenta, and the orientation of filaments and walls in the cosmic web.

Note that this does not imply that every individual halo experiences a large spurious effect: at separations $|\Vector{r}|\ll L$, the periodic force approaches the isolated Newtonian force, and the dynamics are approximately isotropic. However, the concern is that in small boxes the cumulative effect of weak, geometry-aligned torques can become large enough to
bias ensemble statistics. Moreover, once the evolution enters the non-linear regime, density and velocity fluctuations are no longer described by independent Fourier modes. Gravity couples modes and scales in the non-linear regime, and power and phase information are exchanged across scales. As a result, a geometry-aligned anisotropy that is seeded on larger scales can propagate to significantly smaller scales via non-linear mode couplings. In physical terms, the large-scale, box-shaped tidal field generated by the periodic images can coherently modulate collapse and accretion, biasing the orientation and internal anisotropy of forming haloes even when their characteristic size is significantly smaller than the linear box size. This motivates our focus on directional and shape-based measurements presented in the next sections.

\section{Simulations}

To study the effects of periodicity in small comoving cubic \Tcube volumes, we run a new suite of cosmological \Nbody simulations with the \GADGETFOUR \citep{2021MNRAS.506.2871S} simulation code, using identical force and mass resolution across the full suite. The main control parameter is the linear box size $L$, which we vary while keeping the particle mass fixed. This ensures that differences in clustering, halo properties, and orientation statistics can be attributed primarily to finite-volume and periodic-kernel effects, rather than to variations in resolution. The simulation suite consists of five box sizes,
$L \in \left\{100,\;50,\;25,\;12.5,\;6.25\right\}\,\mpc$, with particle numbers scaled as $N_{\mathrm{p}} \propto L^3$ in order to maintain a fixed particle mass of $2.2791\cdot10^{7}\,\msol$ with a $1.667\,\kpc$ Plummer-equivalent cubic spline softening \citep{1985A&A...149..135M}. Similarly to the COLIBRE simulations \citep{2026MNRAS.548ag375S}, we adopt the Dark Energy Survey Year 3 \citep[D3A]{2022PhRvD.105b3520A} ‘3$\times$2pt + All Ext.’ \lcdm cosmological parameters in all runs. The adopted cosmology is summarised in Table~\ref{tab:cosmoparams}.

Since our goal is to quantify the combined finite-volume effects relevant for small-box cosmological simulations, we do not attempt to disentangle the impact of missing large-scale modes from that of the anisotropic \Tcube Green function. Instead, we perform average-density simulations that closely resemble the setup of widely used modern simulation suites (e.g.\ COLIBRE or IllustrisTNG), in which both effects are inherently present. We distinguish changes in angle-averaged statistics, to which both effects can contribute, from systematic preferences for orientations relative to the box axes, which probe the reduced rotational symmetry of the periodic setup. A detailed separation of missing-mode effects from the \Tcube suppression of structure formation is beyond the scope of this work and is discussed in \citet{2021MNRAS.503.5638R}.

\begin{table}
\centering
\caption{Cosmological parameters used in all simulations. Values correspond to the DES Year 3 (D3A) ‘3$\times$2pt + All Ext.’ \lcdm cosmology.}
\label{tab:cosmoparams}
\begin{tabular}{l | c}
\hline\hline
Parameter & Value \\
\hline
$\Omega_{\mathrm{m}}$ & 0.306 \\
$\Omega_{\mathrm{b}}$ & 0.0486 \\
$\Omega_{\Lambda}$ & 0.694 \\
$h$ & 0.681 \\
$\sigma_8$ & 0.807 \\
$n_{\mathrm{s}}$ & 0.967 \\
\hline
\end{tabular}
\end{table}

\subsection{Initial conditions}
The initial conditions were generated with the \monofonic \citep{2020ascl.soft08024H, MONOFONIC:3LPT} initial-condition generator, using transfer functions computed with \CAMB \citep{2000ApJ...538..473L, 2011ascl.soft02026L}. All realisations assume Gaussian random initial perturbations.
To minimise discreteness-induced anisotropies associated with the initial particle load, all runs use a gravitational glass configuration \citep{1994astro.ph.10043W} rather than a simple cubic lattice. This choice is important because a simple cubic grid already possesses $\Oh$ symmetry, which might imprint spurious preferred directions onto orientation-sensitive diagnostics \citep{2026PhRvD.113b3510S}.

The initial redshift for all simulations is $z_{\mathrm{init}} = 127$, and second-order Lagrangian perturbation theory (2LPT) was used to generate the initial displacement and velocity fields \citep{2010MNRAS.403.1859J}.

\subsection{Reference volume}
Our reference configuration is a matched pair of $L=100\,\mpc$ boxes with $1200^3$ particles. The initial conditions for this pair were selected from a large ensemble of independent Gaussian realisations such that the average power spectrum closely approximates the ensemble mean, while each individual Fourier mode remains a Gaussian random variable \citep{2019A&A...631A.160H}. This procedure significantly reduces sample variance in the pair-averaged power spectrum.
We note that previous work has shown that even $L\simeq 100\,\mpc$ periodic boxes exhibit box-induced anisotropies at the few-percent level by $z=0$  \citep{2021MNRAS.503.5638R}. Our $L=100\,\mpc$ reference simulations may therefore retain residual finite-volume effects, but can provide a useful baseline for exploring how rapidly these effects grow towards even smaller volumes.

\subsection{Small-volume simulations}
All smaller-box simulations ($L \le 50\,\mpc$) are initialised from independent Gaussian realisations, using the same cosmology and particle mass as in the $L=100\,\mpc$ reference. This design allows us to characterise the dependence of periodic-box anisotropy on volume while keeping numerical resolution fixed.
As the box size decreases, cosmic variance increases and fewer resolved haloes are available, weakening statistical constraints. In typical simulation suites with only one or a few realisations per box size, the total sampled volume also decreases, allowing increasing statistical fluctuations to obscure (even increasing) systematic differences. In this work, we compensate by deliberately increasing the number of independent realisations towards smaller box sizes, reducing the uncertainty in the ensemble mean, and revealing systematic effects that would be difficult to unambiguously identify in individual volumes.
An overview of the simulation suite is provided in Table~\ref{tab:simulations}.

\begin{table}
\centering
\caption{Overview of the simulation suite. All runs use identical particle mass, periodic boundary conditions, and gravitational glass initial conditions. The $L=100\,\mpc$ configuration consists of a matched pair of realisations with reduced cosmic variance.}
\label{tab:simulations}
\begin{tabular}{l | ccc}
\hline\hline
Run & $L$ [$\mpc$] & $N_{\mathrm{p}}$ & \# realisations\\
\hline
N1200 & 100.0 & $1200^3$ & 2 \\
N600  & 50.0  & $600^3$  & 16 \\
N300  & 25.0  & $300^3$  & 96 \\
N150  & 12.5 & $150^3$  & 128 \\
N75   & 6.25 & $75^3$   & 256 \\
\hline
\end{tabular}
\end{table}

\subsection{Simulation outputs}

During each simulation, we stored a total of 37 particle snapshots, together with halo and subhalo catalogues, covering the redshift range $41 \geq z \geq 0$. Haloes were identified using the built-in friends-of-friends (FOF) halo finder of \GADGETFOUR, with a standard linking length of $b=0.2$ times the mean inter-particle separation. Substructures were identified on-the-fly using the \SUBFIND algorithm \citep{2001MNRAS.328..726S}, which decomposes each FOF halo into gravitationally bound subhaloes.

To investigate how periodicity affects assembly histories, we constructed merger trees for all subhaloes using the built-in GADGET4 post-processing tools. These merger trees enable us to trace the temporal evolution of subhalo properties and to quantify potential geometry-induced biases during structure formation.

\section{Results}

\begin{figure*}[h!]
    \centering
    \begin{subfigure}[a]{0.95\textwidth}
        \centering
        \includegraphics[width=\linewidth]{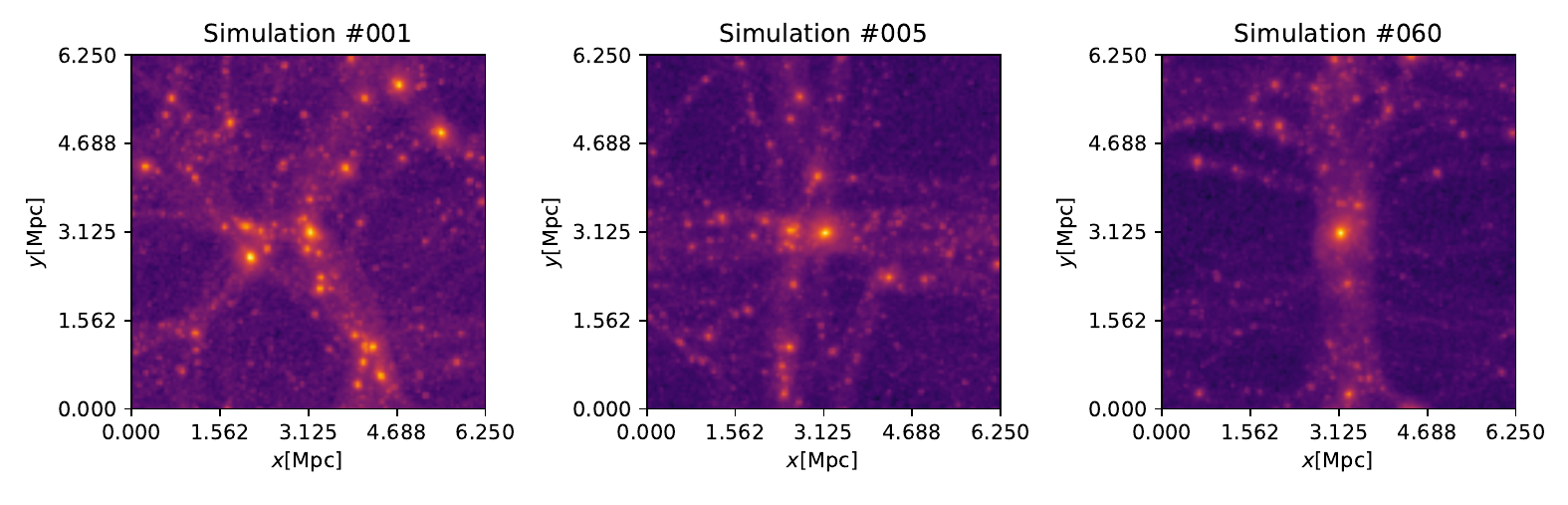}
        \caption{$L=6.25\,\mpc$}
    \end{subfigure}
    ~
    \begin{subfigure}[a]{0.95\textwidth}
        \centering
        \includegraphics[width=\linewidth]{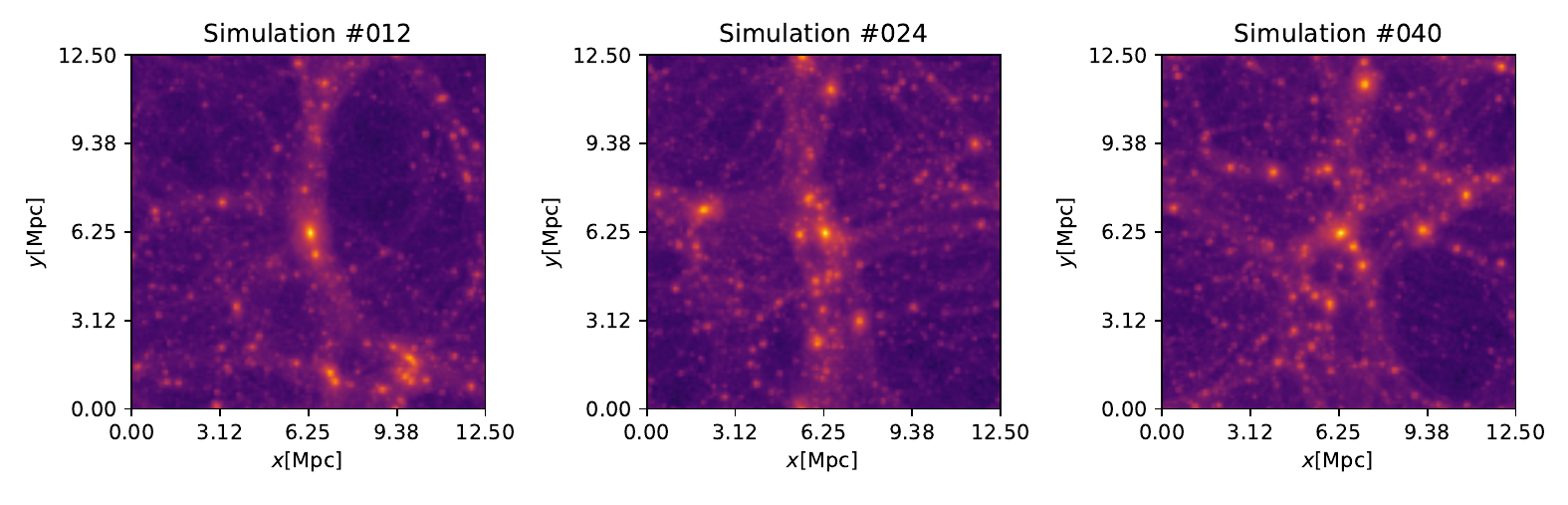}
        \caption{$L=12.5\,\mpc$}
    \end{subfigure}
    ~
    \begin{subfigure}[a]{0.95\textwidth}
        \centering
        \includegraphics[width=\linewidth]{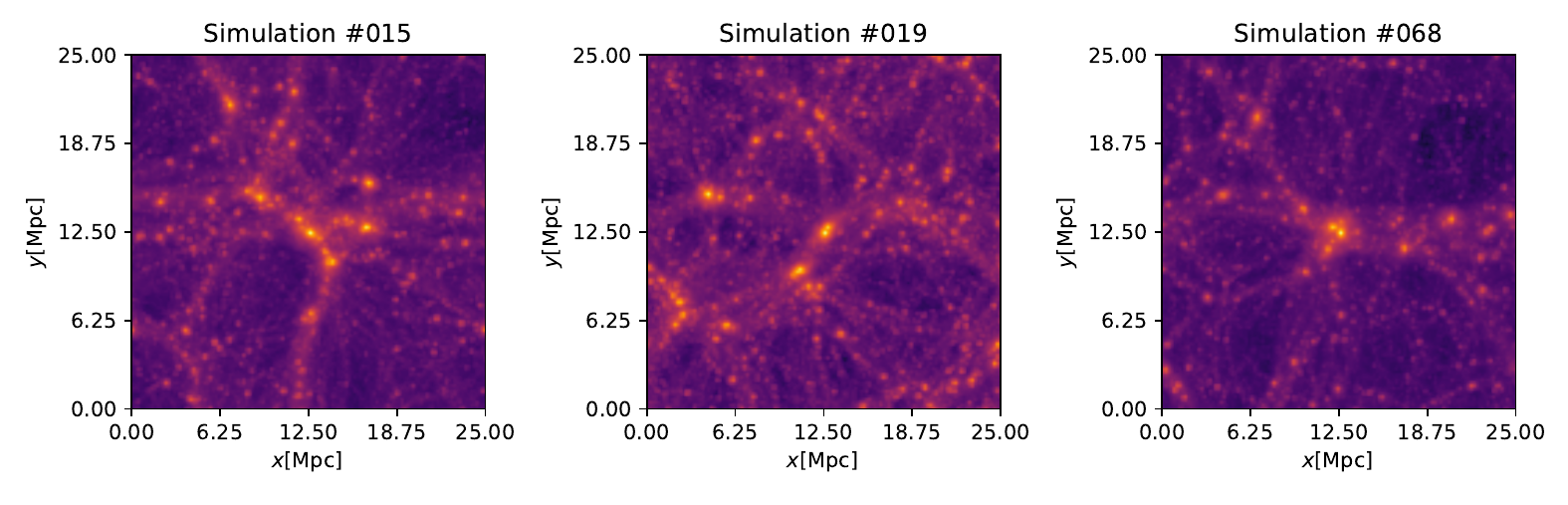}
        \caption{$L=25.0\,\mpc$}
    \end{subfigure}
    \caption{Visualisation of the projected matter density field in randomly selected small-volume boxes at redshift $0$. By using the translation invariance of the periodic \Tcube volume, we shifted the most massive halo to the $[L/2,\,L/2,\,L/2]$ coordinates for an easier comparison. It is clearly visible that most of the formed filaments in these volumes are closely aligned with the face directions of the box.}
    \label{fig:dens_proj_random}
\end{figure*}

\subsection{Matter density field}

\begin{figure*}[h!]
    \centering
    \begin{subfigure}[a]{0.475\textwidth}
        \centering
        \includegraphics[width=\linewidth]{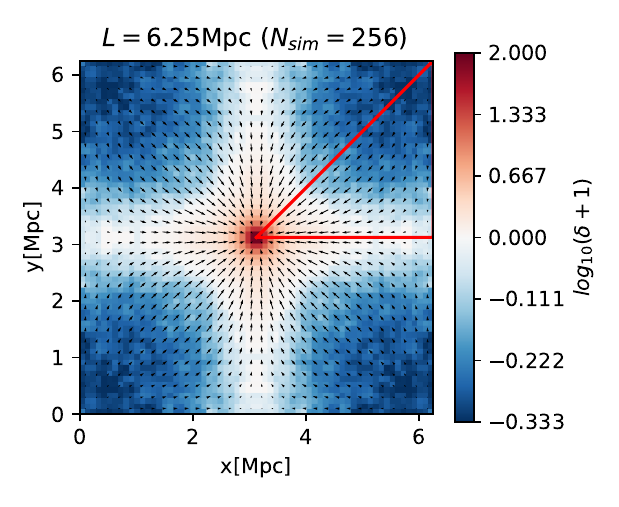}
        \caption{$N75$ simulations}
    \end{subfigure}
    ~
    \begin{subfigure}[a]{0.475\textwidth}
        \centering
        \includegraphics[width=\linewidth]{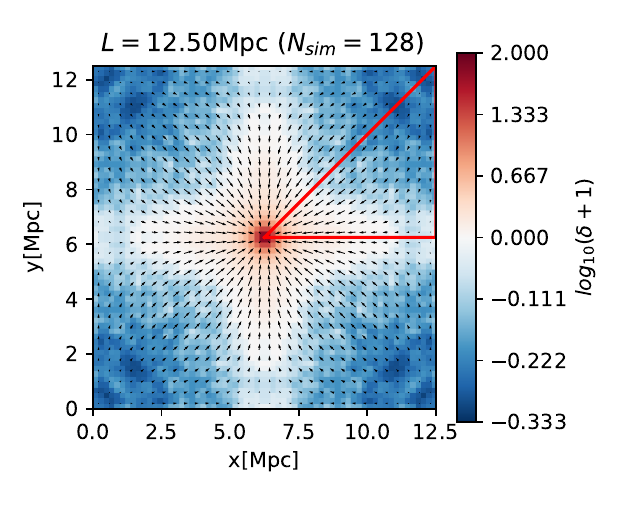}
        \caption{$N150$ simulations}
    \end{subfigure}
    ~
    \begin{subfigure}[a]{0.475\textwidth}
        \centering
        \includegraphics[width=\linewidth]{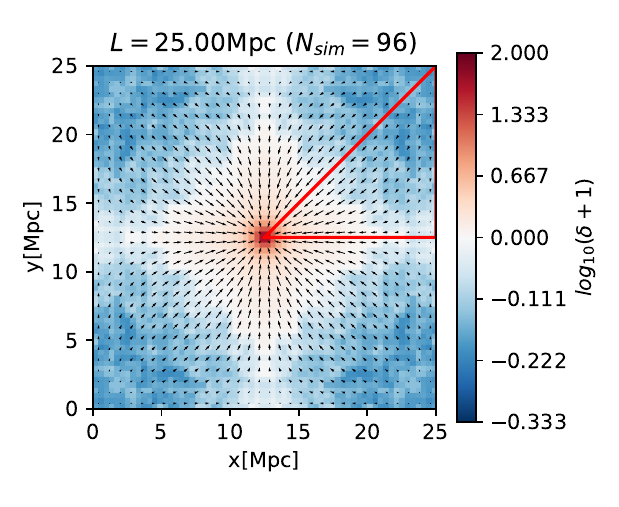}
        \caption{$N300$ simulations}
    \end{subfigure}
    ~
    \begin{subfigure}[a]{0.475\textwidth}
        \centering
        \includegraphics[width=\linewidth]{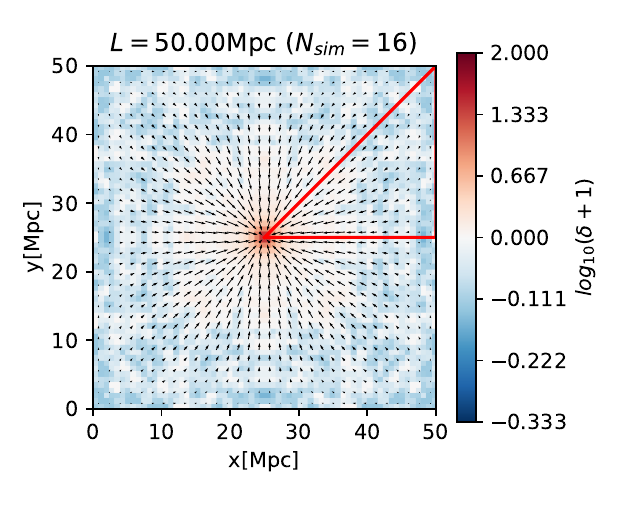}
        \caption{$N600$ simulations}
    \end{subfigure}
    \caption{$\Oh$-symmetrised and realisation-averaged density and velocity fields at $z=0$. For each box size $L$, we centre each realisation on the most massive halo and apply the $\Oh$ symmetry operations of the cube to fold the field into the fundamental domain before averaging. \textbf{Colour map:} $\Oh$-symmetrised mean projected $\delta = \rho/\overline{\rho} - 1$ overdensity field in the $x$--$y$ plane. \textbf{Vectors:} $\Oh$-symmetrised mean peculiar velocity field in the same projection.}
    \label{fig:dens_vel_symm}
\end{figure*}

We begin by examining the signatures of periodic-box anisotropy in the simulated $\delta(\Vector{x}) = \rho(\Vector{x})/\overline{\rho}-1$ overdensity field. Since the three-torus \Tcube is translationally invariant, we apply a global translation to each realisation such that the most massive halo at $z=0$ is placed at the box centre, $\Vector{x}_{\mathrm{c}}=(L/2,\,L/2,\,L/2)$. 

Figure~\ref{fig:dens_proj_random} shows randomly selected small-volume realisations at $z=0$, visualised as projected matter density in the $x$--$y$ plane after centring on the most massive halo. In the smallest boxes, the cosmic web morphology is clearly not isotropic, and the dominant filaments tend to align with the symmetry axes of the cubic simulation volume. This qualitative impression is consistent with a geometry-induced anisotropy that selects the face-normal and edge directions of the periodic domain. To sharpen these signatures, we exploit the discrete $\Oh$ symmetry of the cubic volume (Sect.~\ref{sec:symmetries}) and fold each centred realisation into the fundamental domain and average the resulting symmetrised fields over all realisations at fixed $L$.

The resulting $\Oh$-symmetrised and realisation-averaged overdensity fields are shown in Fig.~\ref{fig:dens_vel_symm}. A prominent, box-aligned ``cross'' pattern emerges around the central halo, with enhanced density extending preferentially along the coordinate axes and suppressed density in between.
In projection, the ensemble-averaged density field in a small \Tcube forms a cross-like pattern of filaments aligned with the box axes and meeting at the most massive halo, with underdense regions between them.
Overlaid on the same panels, as expected, the ensemble-averaged velocity field exhibits coherent inflow towards the central halo. Importantly, however, this inflow is also anisotropic, rather than being spherically symmetric as in \Rcube space: the velocity convergence is strongest along the body diagonals selected by the cubic geometry.

To quantify the box-aligned anisotropy visible in Fig.~\ref{fig:dens_vel_symm}, we define the axis--diagonal contrast (ADC) as
\begin{equation}
    ADC(L,\theta_c) = \langle \delta(L) \rangle_{\theta_c,axis} - \langle \delta(L) \rangle_{\theta_c, diagonal}
    \label{eq:ADC}
\end{equation}
where $\langle \delta(L) \rangle_{\theta_c,\mathrm{axis}}$ denotes the mean value of the $\Oh$-symmetrised overdensity field inside a cone of half-opening angle $\theta_c$ oriented towards the fundamental axis direction, and $\langle \delta(L) \rangle_{\theta_c,\mathrm{diagonal}}$ denotes the corresponding mean inside a cone oriented towards the fundamental body-diagonal direction. Both cones are restricted to the same sphere of radius $L/2$, centred on the reference halo, ensuring identical radial limits for the axis and diagonal measurements.

\begin{figure*}[ht!]
    \centering
    {\includegraphics[width=0.99\textwidth]{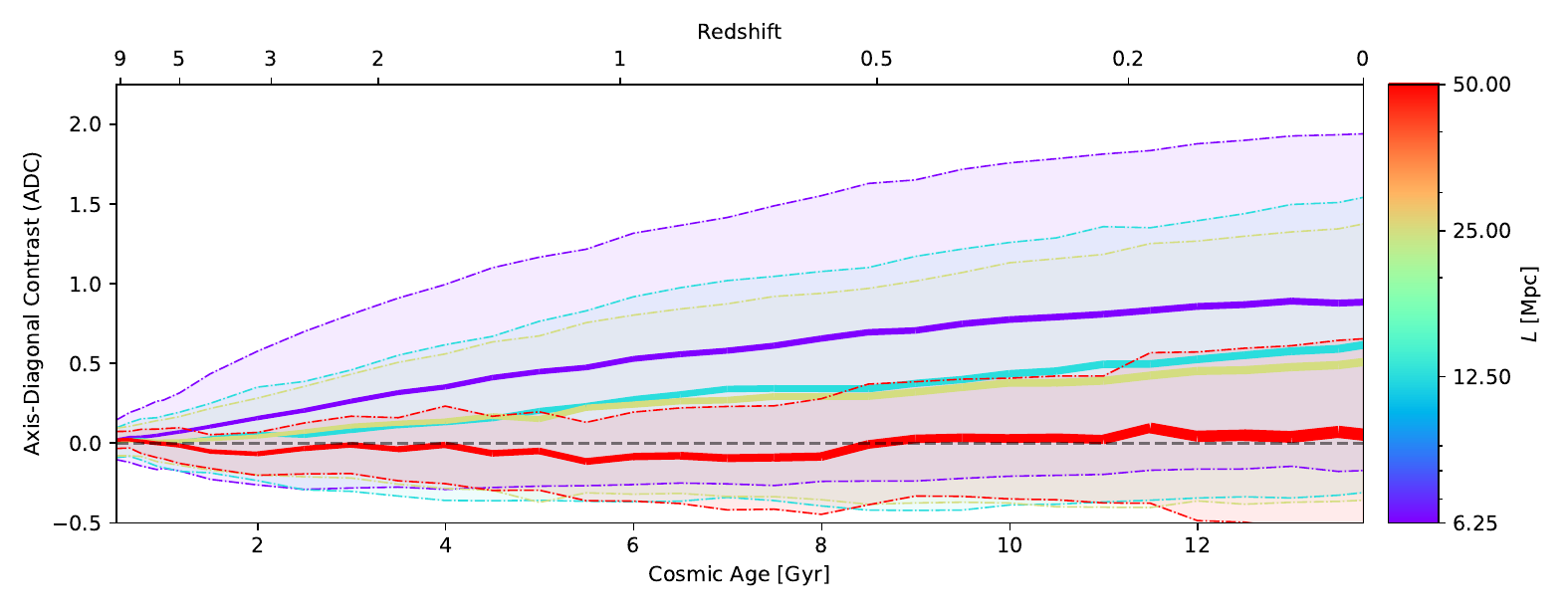}}
     \caption{Density-field anisotropy measured within a sphere of radius $L/2$, as a function of cosmic age (bottom axis) and redshift (top axis). Solid lines represent the ensemble mean Axis-Diagonal Contrast (ADC), while the dash-dotted lines denote the $1\sigma$ cosmic variance between individual simulation volumes. Curves are colour-coded by the linear box size ($L_{box}$). The positive mean ADC demonstrates how the periodic \Tcube topological manifold explicitly breaks the standard cosmological assumption of isotropy. The error of the mean is visualised as the thickness of the solid lines, and this error is significantly smaller than the cosmic variance scatter.}
      \label{fig:dens_ADC}
\end{figure*}

Because the field has already been symmetrised under the full $\Oh$ group, these two cones can be evaluated entirely within the fundamental tetrahedron, which contains all independent information. Positive values of $ADC$ therefore indicate that the symmetrised density field is enhanced along the box axes relative to the diagonal directions, as expected for the box-aligned ``cross'' pattern seen in Fig.~\ref{fig:dens_vel_symm}. We calculated this quantity for all independent realisations with $\theta_c=20^\circ$, and plotted the mean ADC with the measured cosmic variance on Fig.~\ref{fig:dens_ADC}. At early times ($z \gtrsim 5$), the mean contrast across all simulation suites remains close to zero, consistent with an initially isotropic field. However, as structure evolves, the ensemble-mean $\mathrm{ADC}$ grows monotonically, most pronounced in the smallest boxes, indicating the progressive emergence of coherent box-aligned anisotropy. We note a clear dependence on the simulation volume, where smaller boxes experience a significantly more rapid and pronounced growth in directional contrast. This measured anisotropy is highly significant on average and cannot be attributed to statistical noise. 

\subsection{Power spectra}

For a statistically isotropic field, the 1D power spectrum $P(k)$ is defined as
\begin{equation}
\left\langle \delta(\vec{k})\,\delta^*(\vec{k}') \right\rangle
= (2\pi)^3\,\delta_{\rm D}(\vec{k}-\vec{k}')\,P(k),
\qquad k\equiv|\vec{k}|,
\label{eq:Pk_def}
\end{equation}
where $\delta_{\rm D}$ is the Dirac delta function and 
\begin{equation}
\delta(\vec{k}) = \int \delta(\vec{x})\,e^{-i\vec{k}\cdot\vec{x}}\,{\rm d}^3x
\label{eq:delta_k_continuous}
\end{equation}
is the 3D Fourier transform of the matter overdensity of the simulation. This quantity describes how cosmic matter is distributed across various scales.

Missing large-scale modes and periodic gravitational interactions can suppress structure growth in small boxes \citep{2021MNRAS.503.5638R}. The matter power spectrum allows us to quantify how this suppression depends on scale, redshift, and box size.

It is worth mentioning that the Separate Universe (SU) simulation formalism \citep{2015MNRAS.448L..11W, 2018MNRAS.479..162S, 2019MNRAS.488.2079B} can be used to include the effects of the missing modes, albeit at the cost of running multiple simulations with different mean densities. More importantly, this method is restricted to larger boxes ($L \simeq 100\mpc h^{-1}$ for $z=0$), since the unresolved modes are treated as linear density perturbations with infinite wavelength. It cannot account for mode couplings from intermediate scales (e.g., modes that are larger than $L$ but smaller than the super-survey mode).

\begin{figure*}[ht!]
    \centering
    {\includegraphics[width=0.99\textwidth]{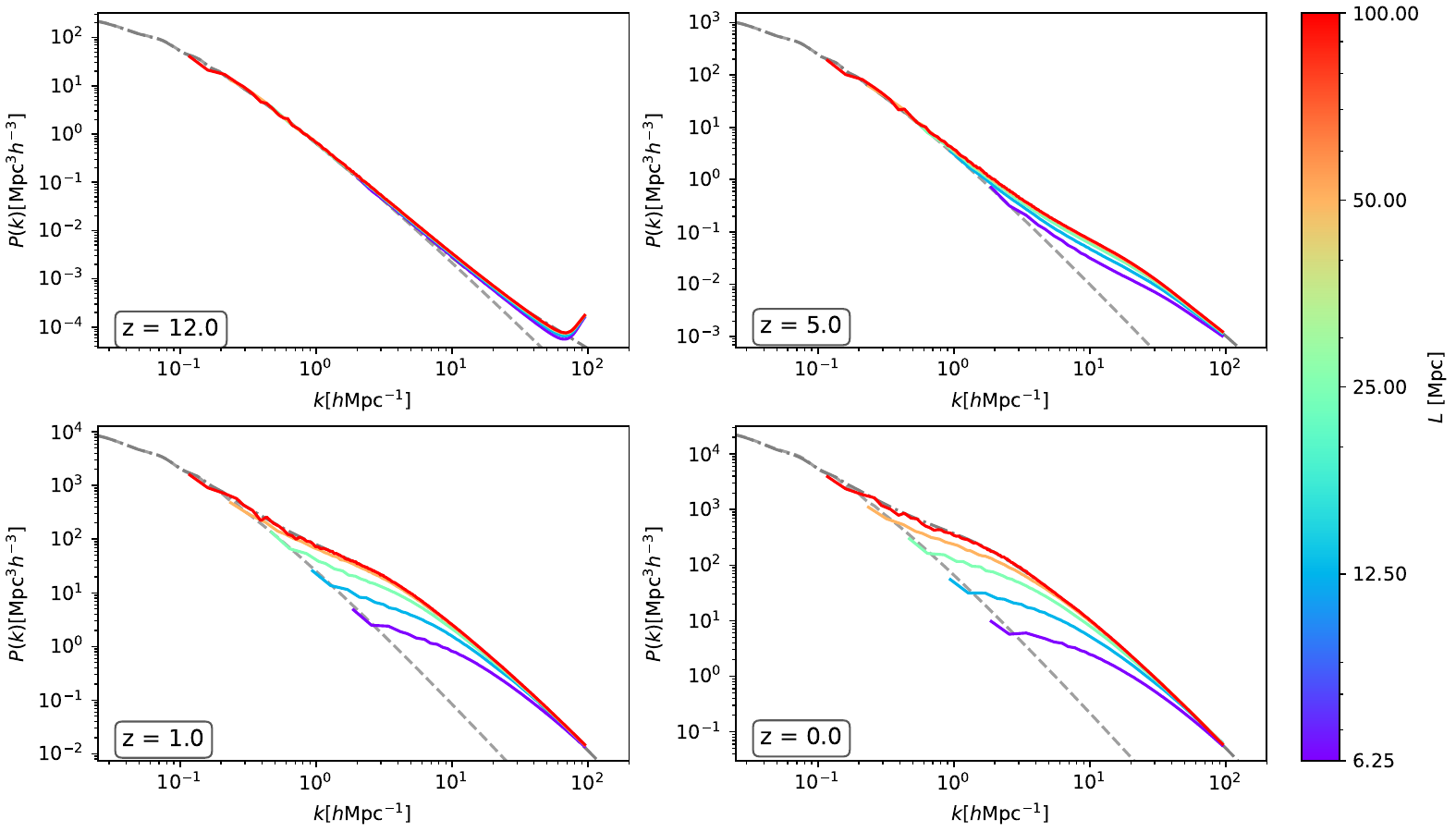}}
     \caption{Ensemble average of the matter power spectrum at different redshifts and simulation box sizes. The grey dot-dashed line represents the expected non-linear power spectrum calculated by \HMCODE \citep{2021MNRAS.502.1401M}, while the grey dashed line shows the \CAMB linear power spectrum. If the fundamental mode reaches the non-linear regime, power is strongly suppressed on all scales due to finite-volume effects.}
      \label{fig:meanPk}
\end{figure*}

We calculated the power spectrum of all simulations in our suite with \nbodykit \citep{Hand_2018} using a triangular-shaped-cloud density-assignment function on grids of size $N_{\rm grid} \in \{2048, 1024, 512, 256, 128\}$ for $L \in \left\{100,\;50,\;25,\;12.5,\;6.25\right\}\,\mpc$ box sizes, respectively. We applied deconvolution of this window in Fourier space and we used interlacing \citep{2016MNRAS.460.3624S} to reduce aliasing. The calculated ensemble mean power spectra for all box sizes at four selected redshifts are plotted in Fig.~\ref{fig:meanPk}. As a consequence of the missing modes and the weaker gravity in \Tcube, the power spectrum we measured is significantly suppressed in our $L=6.25\mpc$ and $L=12.5\mpc$ simulations even at $z=5$. At redshift $1.0$, this becomes true for all simulation suites with $L\leq 25\mpc$. By $z=0$, the smallest boxes are orders of magnitude below the non-linear prediction, and even the largest $L=100\,\mpc$ pair shows a percent-level deficit.

\subsection{Halo abundances and formation history}

\begin{figure*}[ht!]
    \centering
    {\includegraphics[width=0.99\textwidth]{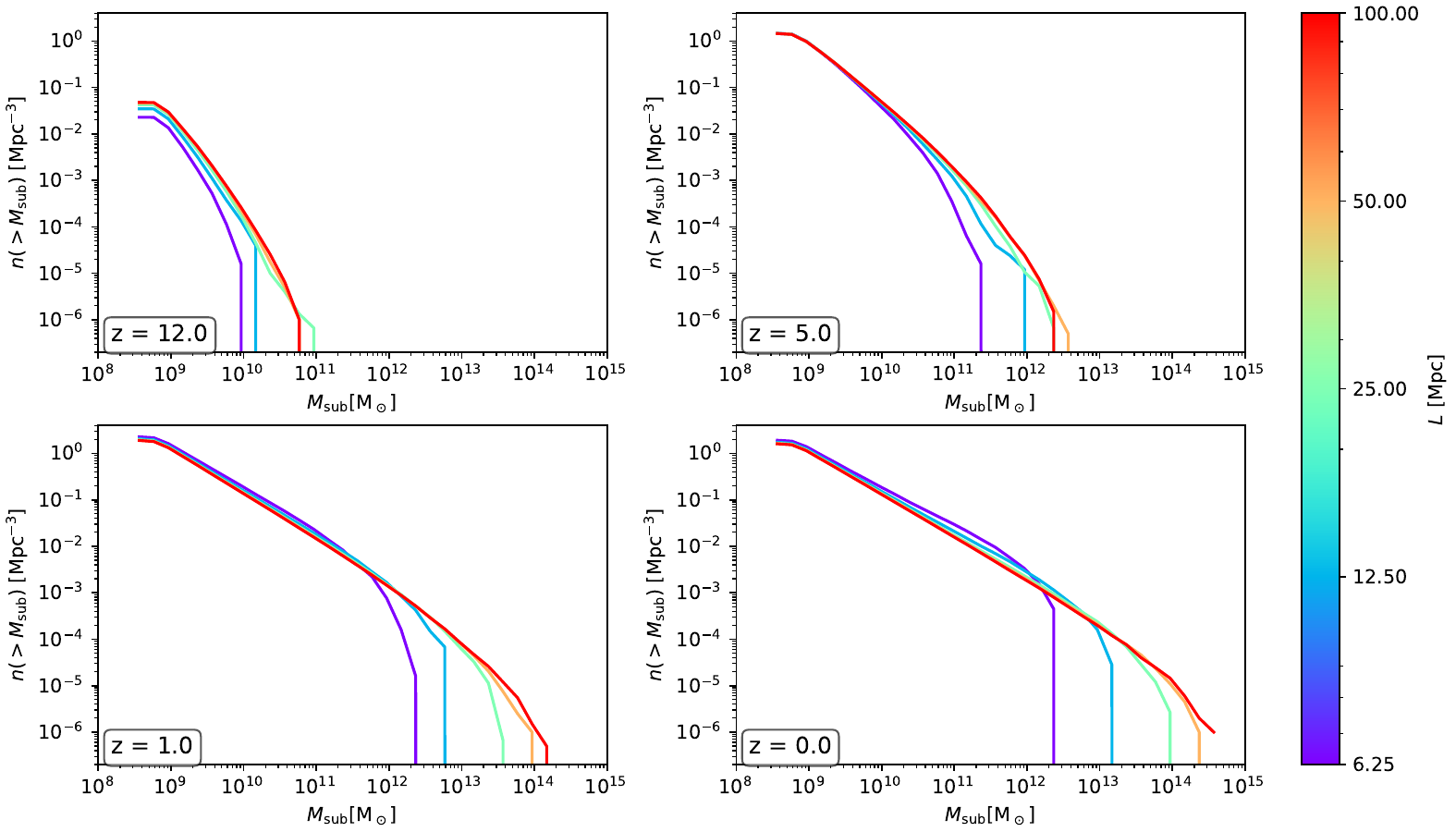}}
    \caption{The cumulative subhalo mass function $n(>M_{\mathrm{sub}})$ at different redshifts and in different periodic simulation volumes. We used all realisations at each box size to reconstruct this ensemble average. While the halo number density monotonically increases with the box size at high redshift in all mass bins, this changes dramatically at low redshifts. At late times, the lower-mass haloes are more abundant in the smaller boxes than in the reference $L\,=\,100\,\mpc$ simulation pair.}
      \label{fig:sHMF}
\end{figure*}

In dark-matter-only simulations, subhaloes serve as the standard tracers of galaxy hosts and satellites. Therefore, if the abundance, formation history, or merger statistics of subhaloes are biased by periodic-box effects, one should expect corresponding biases in the predicted population and evolutionary histories of galaxies in hydrodynamical simulations performed in the same volumes.

We first examine the cumulative subhalo mass function, $n(>M_{\mathrm{sub}})$, as a function of redshift and box size. Since the particle mass is fixed across the full suite, the same subhalo mass threshold corresponds to the same effective resolution in every simulation, allowing a direct comparison between different volumes. The ensemble-averaged cumulative subhalo mass functions at four redshifts are shown in Fig.~\ref{fig:sHMF}. Differences in the halo mass function are apparent at all redshifts. At early times, smaller boxes have lower subhalo number densities above a given mass threshold. At late times, this trend reverses at low masses, where smaller boxes have higher cumulative abundances than the reference $L=100\,\mpc$ simulation pair.

\begin{figure*}[ht!]
    \centering
    {\includegraphics[width=0.99\textwidth]{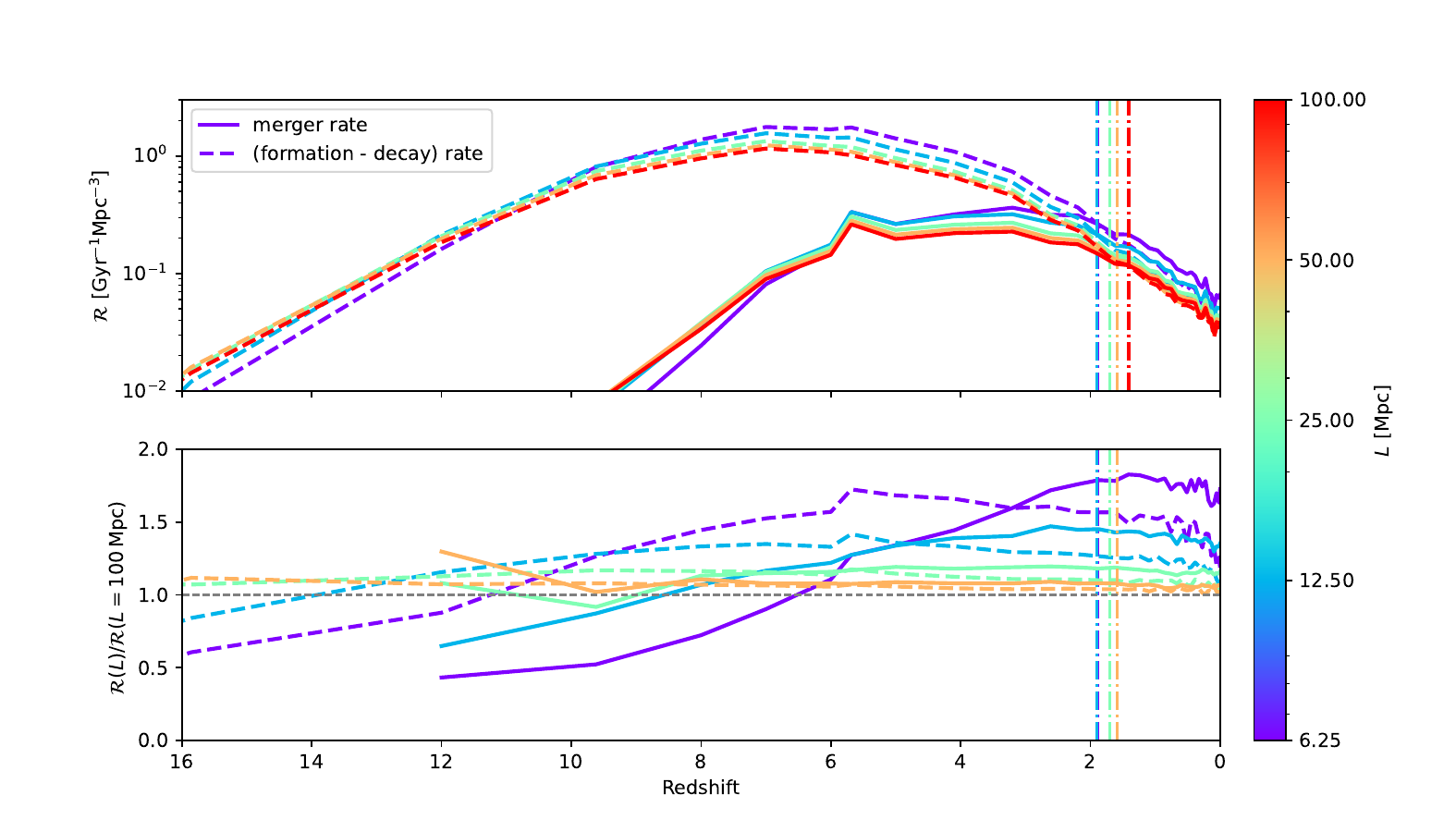}}
    \caption{Ensemble average of the subhalo formation, decay, and merger rates as a function of redshift in different simulation box sizes. The solid lines show the total $\mathrm{merger}$ rate, while the dashed lines represent the $(\mathrm{formation} - \mathrm{decay})$ rates. \textbf{Top:} Rates per unit cosmic time and comoving volume, in $\mathrm{Gyr}^{-1}\,\mpc^{-3}$. Vertical lines correspond to the redshift at which these two quantities are equal. After this point, the total number of subhaloes in the simulation decreases. \textbf{Bottom:} Ratios of the net formation rate (formation minus decay) and the merger rate to their respective values in the $L=100\,\mpc$ reference simulation pair. Both the $(\mathrm{formation} - \mathrm{decay})$ and $\mathrm{merger}$ rates converge to their reference-box values as $L$ increases, but even our $L=50\,\mpc$ suite shows a $\sim 6\%$ increase in the merger rate at $z=0$. }
      \label{fig:subhaloMergerRates}
\end{figure*}

To characterise the temporal evolution of the subhalo population in more detail, we reconstruct the subhalo assembly history from the merger trees and measure three rates per comoving volume and per unit cosmic time: the subhalo formation rate, the merger rate, and the decay rate. 
Here the formation rate counts newly appearing subhaloes in the catalogue, the merger rate counts subhalo--subhalo mergers, and the decay rate counts subhaloes that disappear from the catalogue, for example because of tidal disruption or stripping below the resolution threshold. In practice, the formation and merger terms dominate the net evolution and are therefore the most informative quantities to compare across box sizes. These quantities as a function of redshift are plotted in Fig.~\ref{fig:subhaloMergerRates}. It is clear that the subhalo merger and formation rates depend strongly on the linear box size of the \Tcube simulation: both the merger and formation rates increase as the box size decreases after redshift~6. The transition redshift $z_{\rm turn}$, at which the merger rate exceeds the net formation rate (formation minus decay) and the total subhalo abundance begins to decline, also depends strongly on box size: it starts from $z_{\rm turn} = 1.89$ for box sizes $L\,=\,6.25\,\mpc$, reaches $z_{\rm turn} = 1.59$ for boxes with $L\,=\,50\,\mpc$, while in the reference $L\,=\,100\,\mpc$ simulation pair $z_{\rm turn}$ is $1.40$. The significantly different merger history between the simulations implies different subhalo population statistics and parameters between the different box sizes at the same redshifts.

\subsection{Halo shapes and orientations}

\begin{figure*}[ht!]
    \centering
    {\includegraphics[width=0.99\textwidth]{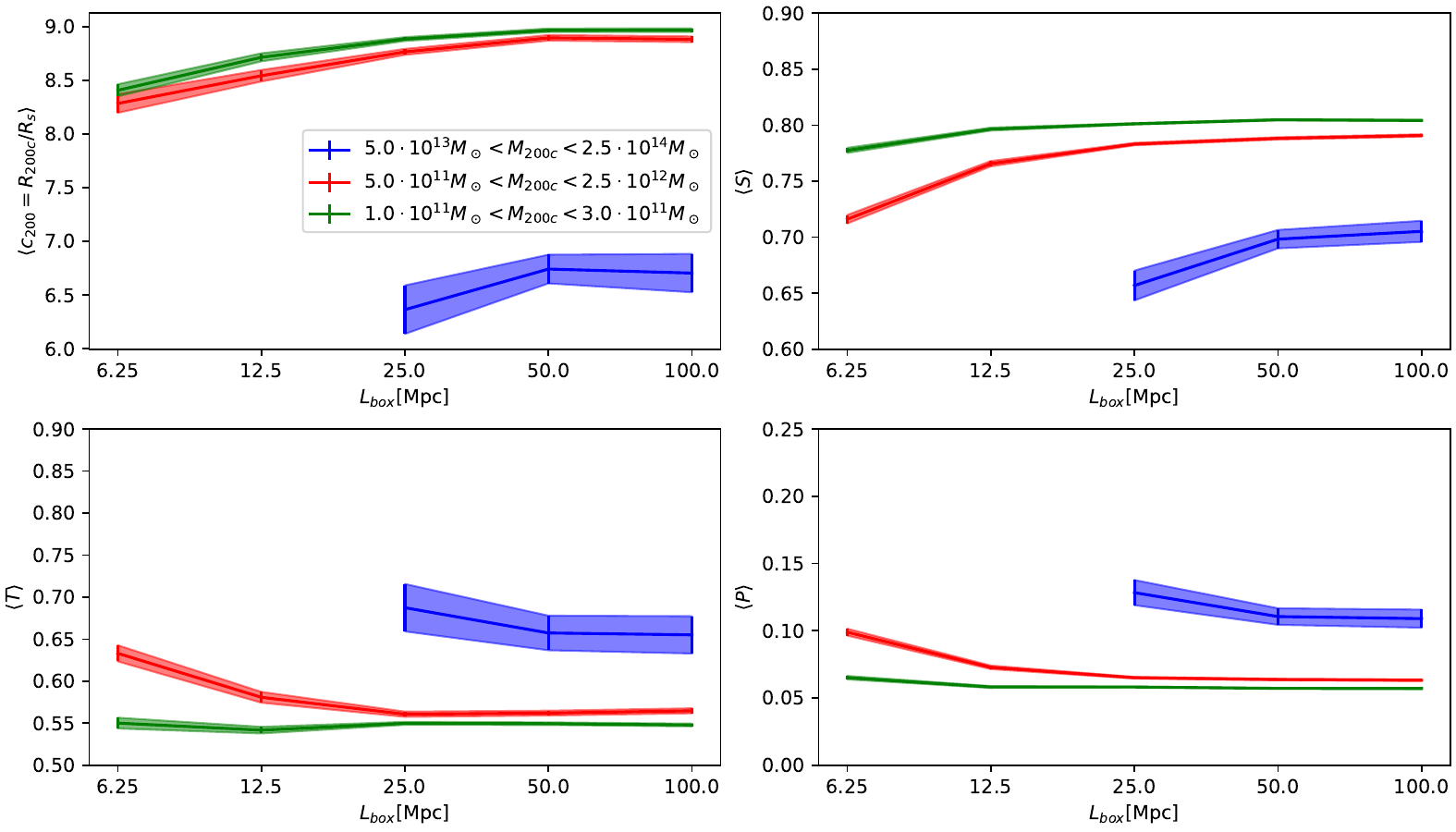}}
    \caption{Ensemble average of the FoF halo concentration and shape parameters as a function of the linear box size $L$ at $z=0$ for LMC-, MW-, and M87-mass dark matter haloes. The error bars represent the measured error of the mean of these shape parameters. It is clear that the mean concentration and halo shape parameters depend on the simulation box size, and small boxes are strongly biased at redshift 0.}
      \label{fig:haloeshapeParams}
\end{figure*}

To quantify the impact of the periodic box size on the structural properties of dark matter haloes at the final state, we analyse their concentration, sphericity, triaxiality, and prolateness, along with their orientation. We select distinct halo samples across three characteristic mass bins: a Large Magellanic Cloud (LMC)-like bin ($1\times10^{11} \msol< M_{\text{200c}} < 3\times10^{11} \, \msol$), a Milky Way (MW)-like bin ($5\times10^{11} \msol< M_{\text{200c}} < 2.5\times10^{12} \, \msol$), and an M87-like bin ($5\times10^{13} \msol< M_{\text{200c}} < 2.5\times10^{14} \, \msol$). Note that no haloes in the latter mass range are present in the $L=6.25\,\mpc$ and $L=12.5\,\mpc$ suites.

Using the FoF catalogues at redshift~0, we identify isolated central haloes within the designated mass bins, extracting their centre-of-mass coordinates and $R_{200c}$ radii. For each selected halo, we extract all dark matter particles within a spherical volume of radius $1.125 \times R_{200c}$ from the corresponding particle snapshot. To isolate the smooth halo component, particles belonging to all other subhaloes identified by \SUBFIND within this volume are removed prior to analysis \citep{2011ApJS..197...30Z}. The shape of each halo is determined using the unreduced shape tensor (often called the inertia tensor), 
\begin{equation}
    I_{ij} = \sum_{k=1}^{N_p} x_{k,i} x_{k,j}
    \label{eq:momentofinertiatensor}
\end{equation}
where the sum runs over the $N_p$ retained dark matter particles within $1.125 R_{200c}$, and $x_{k,i}$ denotes the particle coordinates relative to the halo centre of mass. Diagonalising $I_{ij}$ yields the principal-axis eigenvalues, sorted such that $\lambda_1 \ge \lambda_2 \ge \lambda_3$. We define axis measures $a=\sqrt{\lambda_1}$, $b=\sqrt{\lambda_2}$, and $c=\sqrt{\lambda_3}$, corresponding to the major, intermediate, and minor axes. Their common normalisation cancels in the dimensionless shape parameters below. From these axes, we compute three fundamental morphological indicators: sphericity ($S$), triaxiality ($T$), and prolateness ($P$), as follows:

\begin{equation}
    S = \frac{c}{a}, \qquad T = \frac{a^2 - b^2}{a^2 - c^2}, \qquad P = \frac{a - b}{a + b}.
    \label{eq:morphology}
\end{equation}

By calculating these shape parameters for all haloes in our mass bins across our suite of simulations, we compute the ensemble averages ($\langle S \rangle$, $\langle T \rangle$, and $\langle P \rangle$) as a function of the periodic box size $L$ in Fig.~\ref{fig:haloeshapeParams} at redshift 0. We observe a clear, systematic dependency of all three shape quantities on the box size $L$. This trend indicates that the finite-volume effects dictated by the \Tcube topology not only influence the large-scale matter distribution, power spectrum, and halo mass distributions, but directly propagate down to the internal halo properties.

The eigenvectors of $I_{ij}$ corresponding to the major, intermediate, and minor axes are reduced to unit vectors $\hat{\mathbf{e}}_a$, $\hat{\mathbf{e}}_b$, and $\hat{\mathbf{e}}_c$. We map every orientation unit vector of the MW-like haloes ($5\times10^{11} \msol< M_{\text{200c}} < 2.5\times10^{12} \, \msol$) into the fundamental tetrahedron of the cube using the transformation rules established in Section \ref{sec:symmetries}. To quantify the deviation from isotropy within this fundamental wedge, we apply a two-sample Kolmogorov-Smirnov (K-S) test to the dominant spatial components of the folded vectors. After mapping the orientation unit vectors into the fundamental wedge, we extract the maximum component ($x$), which acts as a direct proxy for the degree of alignment with the nearest simulation box axis. We compare this empirical distribution against a reference sample of $10^6$ unit vectors drawn uniformly on the sphere and folded under the same $\Oh$ symmetry. The K-S test evaluates the maximum distance between the cumulative distribution functions of the halo orientations and the isotropic baseline, yielding a $p$-value that we translate into a Gaussian-equivalent significance ($\sigma$) to establish the presence of box-aligned anisotropy. At each box size, our halo samples combine many independent simulation realisations. The quoted K–S significances nevertheless assume independent halo orientations; residual correlations within individual boxes may affect their calibration.

\begin{figure*}[ht!]
    \centering
    {\includegraphics[width=0.99\textwidth]{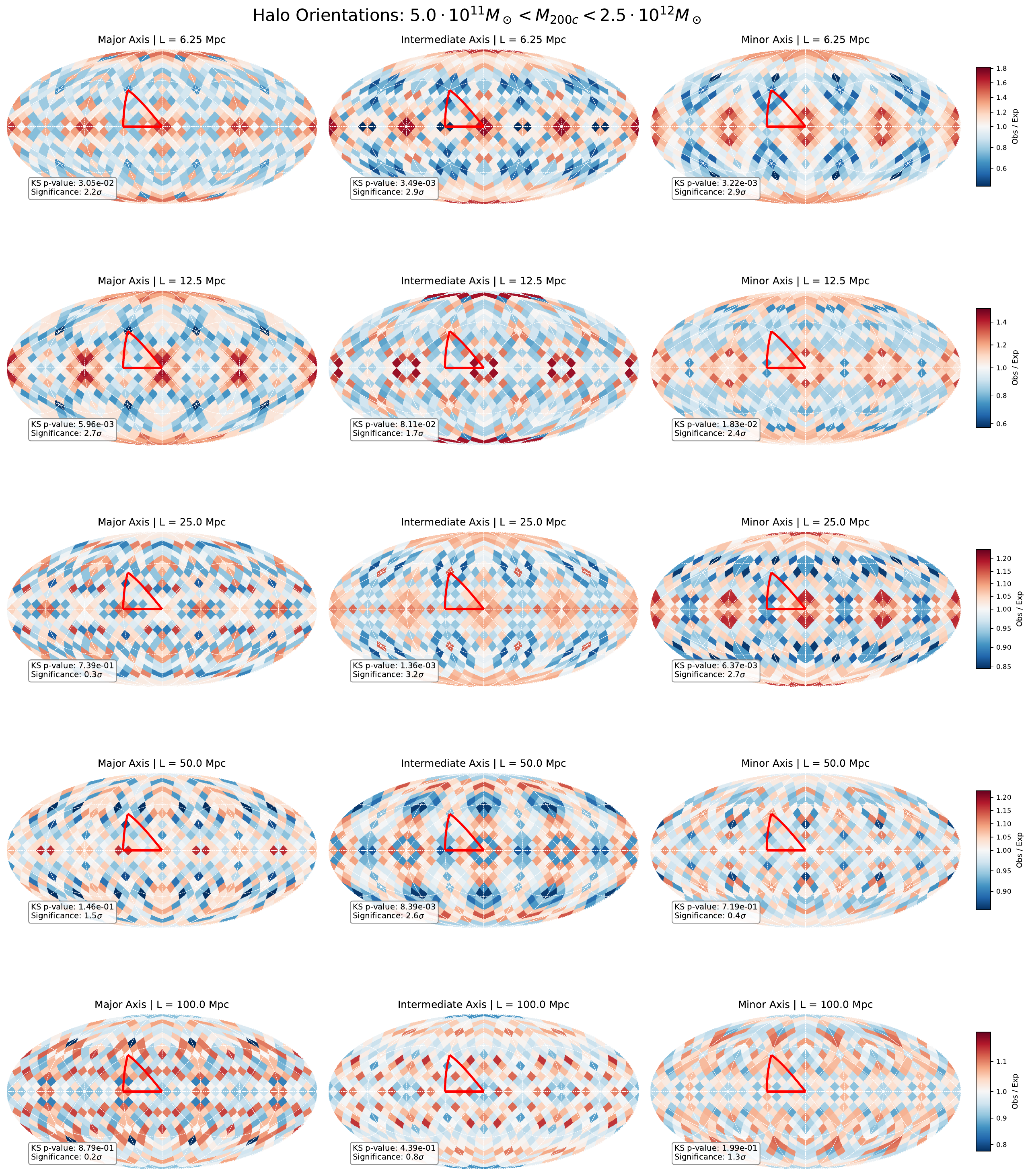}}
    \caption{The distribution of the major, intermediate, and minor axes of all Milky Way-mass ($5\times10^{11} \msol< M_{\text{200c}} < 2.5\times10^{12} \, \msol$) dark matter haloes at redshift 0, in different simulation box sizes. The smaller boxes show significant anisotropy, with the preferred axis directions pointing towards the box faces. Only the largest ($L=100\,\mpc$) simulation pair is fully consistent with statistical isotropy.}
      \label{fig:HaloOrientations_MW}
\end{figure*}

For visualisation, we apply all transformations of $\Oh$ to the folded distribution, producing a symmetrised angular distribution over the full sphere. The final angular probability distributions and the significance of the anisotropy can be seen in Fig.~\ref{fig:HaloOrientations_MW}. The haloes in small boxes display a clear preference to align their axes with the face directions of the \Tcube space. Interestingly, the minor and intermediate axes correlations are much stronger than those of the major axes.

\subsection{Filament orientations}
\begin{figure*}[ht!]
    \centering
    {\includegraphics[width=0.99\textwidth]{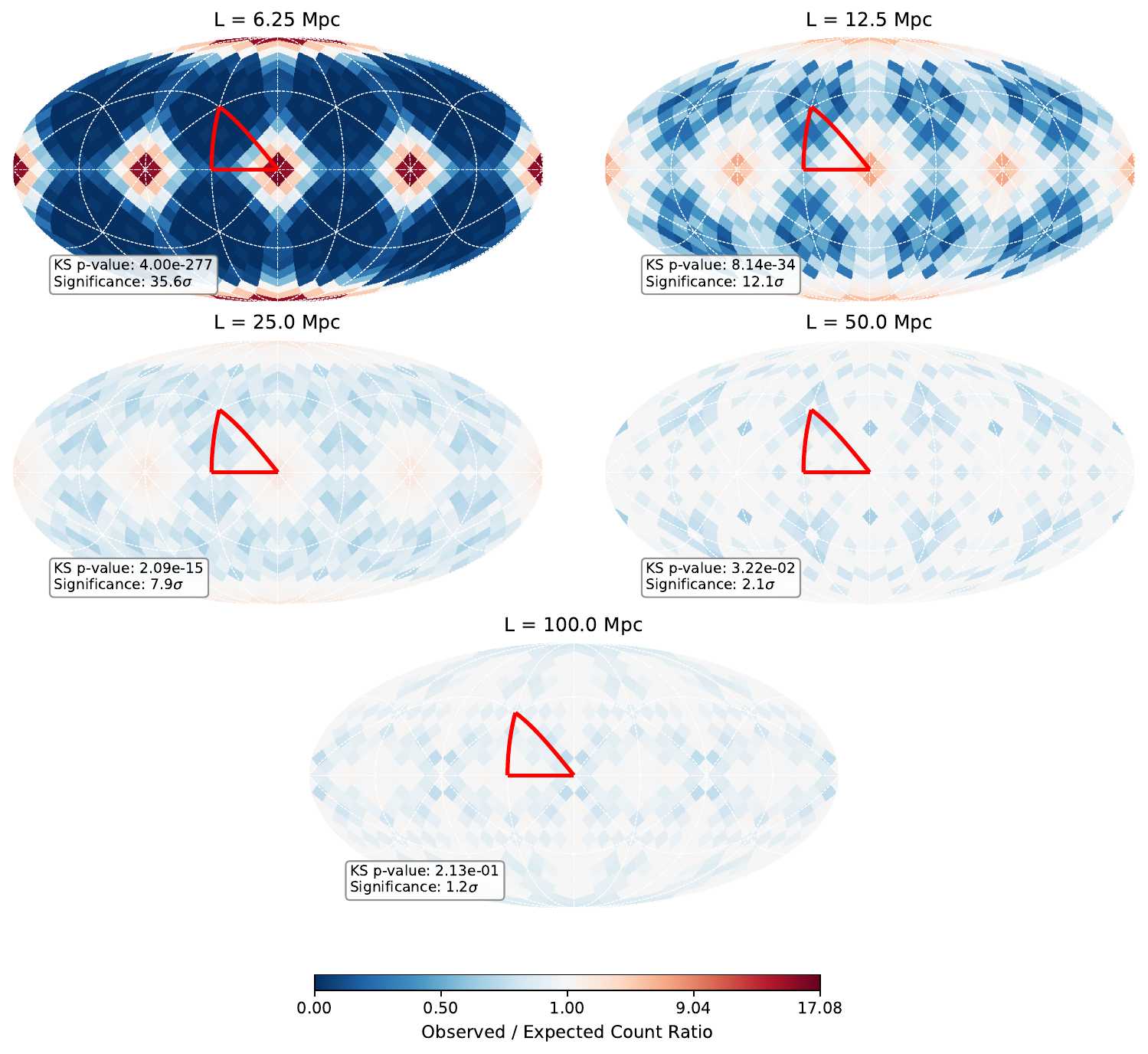}}
    \caption{$\Oh$-symmetrised distribution of the filament orientations in different periodic box sizes at $z=0$.}
      \label{fig:FilamentOrientations}
\end{figure*}

While haloes are affected by the anisotropic gravity of the \Tcube topological manifold, Fig.~\ref{fig:dens_proj_random} and Fig.~\ref{fig:dens_vel_symm} suggest that an even stronger anisotropy signal may be expected for cosmic filaments. To test this, we use the \crispy (Computational Ridge Identification with SCMS for Python) \citep{2020ApJ...891...84C} implementation of the Subspace Constrained Mean Shift (SCMS) algorithm \citep{JMLR:v12:ozertem11a}, and identify the 3D density ridges (filaments) in our FoF dark matter halo catalogues at redshift 0. To determine the primary direction of each filament, we apply principal component analysis (PCA) to its ridge points and take the eigenvector corresponding to the largest eigenvalue as its orientation axis. We then map these axes into the $1/48$ spherical fundamental wedge, analogous to the dark matter halo orientations. The $\Oh$-symmetrised distribution of the filaments can be seen in Fig.~\ref{fig:FilamentOrientations}. Except for our reference simulation pair, the filaments in all small-volume simulations are more likely to be oriented towards the faces of the simulation cube than in any other direction.

To quantify the deviation from isotropy, we employ the same one-dimensional two-sample K-S test introduced for the halo orientations. Considering all filaments across all realisations, we find $\{\text{1.2}, \, \text{2.1}, \, 7.9, \, 12.1, \, 35.6\}\, \sigma$ significance for the filament distribution anisotropy in simulation boxes of $L \in \{100, \, 50, \, 25, \, 12.5, \, 6.25\} \, \text{Mpc}$, respectively. For $L<50\,\mpc$, filament orientations show statistically highly significant departures from isotropy, with a preference for alignment with the simulation box axes. Box-aligned anisotropy is a dominant feature of the cosmic web in small periodic volumes.

\section{Conclusions}

In this work, we investigate the effects of the global \Tcube topological manifold widely used in cosmological simulations. Our cosmological dark-matter-only $N$-body simulations, spanning box sizes from $L = 6.25\,\mathrm{Mpc}$ to $L = 100.0\,\mathrm{Mpc}$, demonstrate that the \Tcube periodic topological manifold acts as an anisotropic tidal field that forces structures to align with the Cartesian axes of the box. Consequently, while extremely small simulation boxes (e.g., $L \sim 12 \, \text{Mpc}$) are frequently employed in modern hydrodynamical simulations to achieve the high resolution required for galaxy formation studies, our findings suggest that these models evolve within a biased macroscopic environment. Through non-linear mode couplings, the box-scale directional preferences propagate down to smaller physical scales, inducing systematic shifts in mean halo shapes, altered subhalo abundances, and modified merger histories. Structural and kinematic statistics derived from such small volumes at late times must be treated with caution. It is expected that these effects and potential biases also propagate to the galaxy populations of small-volume hydrodynamical simulations.

To mitigate these \Tcube topology-induced systematics in cosmological simulations, several practical strategies can be adopted. For constant-resolution runs finished at $z=0$, using comoving box sizes of $L \ge 100\,\mathrm{Mpc}$ ensures that alignment effects drop below the percent level. Alternatively, small-box standalone simulations should be terminated before the scale of non-linear structures grows comparable to the fundamental mode of the box. When high spatial resolution is required for galaxy formation studies, cosmological zoom-in techniques embedded within a large parent volume ($L \ge 100\,\mathrm{Mpc}$) provide the most robust path forward, preserving a realistic large-scale tidal field while avoiding the unphysical artefacts of standalone small periodic domains.

\begin{acknowledgements}
      GR and TS acknowledge support from Research Council of Finland grant 354905.
      The authors thank Oliver Hahn for his help with the initial condition generation software.
      We gratefully acknowledge the use of open source software, including {\sc Astropy} \citep{astropy:2022}, {\sc Numpy} \citep{harris2020array} and {\sc Matplotlib} \cite{Hunter:2007}. All simulations were performed at CSC – IT Center for Science, Finland.
\end{acknowledgements}

\bibliographystyle{aa}
\bibliography{References}

\end{document}